
\normalbaselineskip=12pt
\baselineskip=12pt
\magnification=1200
\hsize 15.0truecm \hoffset 0.5 truecm
\vsize 23.0truecm \voffset 0.5 truecm
\nopagenumbers
\headline={\ifnum \pageno=1 \hfil \else\hss\tenrm\folio\hss\fi}
\pageno=1

\def\lsim{\mathrel{\rlap{\lower4pt\hbox{\hskip1pt$\sim$}}
    \raise1pt\hbox{$<$}}}         
\def\gsim{\mathrel{\rlap{\lower4pt\hbox{\hskip1pt$\sim$}}
    \raise1pt\hbox{$>$}}}         

\def\ea{{\it et al.}}
\def\eg{{\it e.g.}}
\def\ie{{\it i.e.}}
\def\mn{{m_{\scriptscriptstyle N}}}
\def\mz{{M_{\scriptscriptstyle Z}}}
\def\mw{{M_{\scriptscriptstyle W}}}
\def\tw{{\theta_{\scriptscriptstyle W}}}
\def\spw{\scriptscriptstyle W}
\def\spz{\scriptscriptstyle Z}
\def\ga{{g_{\scriptscriptstyle A}}}
\def\gv{{g_{\scriptscriptstyle V}}}
\def\px{{p_{\scriptscriptstyle X}}}
\def\spa{\scriptscriptstyle A}
\def\spv{\scriptscriptstyle V}


\line{\hfil DFTT 80/93}
\line{\hfil December 1993}
\vskip 24pt
\centerline{\bf Polarized deep inelastic scattering at high energies}
\centerline{\bf and parity violating structure functions}
\vskip 36pt\centerline{Mauro Anselmino}
\vskip 6pt
\centerline{\it Dipartimento di Fisica Teorica, Universit\`a di
Torino}
\centerline{\it and Istituto Nazionale di Fisica Nucleare,
Sezione di Torino}
\centerline{\it Via P. Giuria 1, I--10125 Torino, Italy}
\vskip 12pt
\centerline{Paolo Gambino\footnote*{Work partially supported by the
NSF under  grant No. PHY--9017585}}
\vskip 6pt
\centerline{\it Department of Physics, New York University,
4 Washington Place}
\centerline{\it New York, NY 10003, USA}
\vskip 12pt
\centerline{Jan Kalinowski\footnote{**}{Work partially supported by the
Polish Committee for Scientific Research}}
\vskip 6pt
\centerline{\it University of Warsaw, Institute of Theoretical Physics}
\centerline{\it Ul. Hoza 69, 00681 Warsaw, Poland}
\vskip 1.0in
\centerline{\bf ABSTRACT}
\vskip 12pt
\noindent
A comprehensive analysis of deep inelastic scattering of polarized
charged leptons on polarized nucleons is presented; weak interaction
contributions, both in neutral and charged current processes,
are taken into account and the parity violating polarized
nucleon structure functions are studied. Possible ways of their
measurements and their interpretations in the parton model are discussed.
\vfill
\eject
\input harvmac
\baselineskip 12pt plus 1pt minus 1pt
\noindent
{\bf 1 - Introduction}
\vskip 12pt
The nucleon internal structure, as probed in deep inelastic scatterings
(DIS) at high $Q^2$ values, keeps revealing unexpected features and
appears to be much less simple than na\"{\i}vely expected according to
the quark-parton model ideas. The recent experimental data on polarized
structure functions \ref\crisis{J.~Ashman \ea, {\it Phys. Lett.}
{\bf B206} (1988) 364; {\it Nucl. Phys.}, {\bf B328} (1990) 1}
and the Gottfried sum rule \ref\gott{P.~Amaudruz \ea, {\it Phys. Rev.
Lett.} {\bf 66} (1991) 560} have indeed been surprising ones.
In particular, the spin content of protons in terms of its constituents
needs a better and deeper understanding before a clear pattern emerges
and non perturbative effects could be under control; after many
theoretical efforts some more experimental information would certainly
help in casting some light on the subject.

We consider here the deep inelastic scattering of charged
polarized leptons on
polarized nucleons at very high energies, taking into account both
electromagnetic and weak interactions, with both neutral and charged
current contributions. We discuss in detail what can be learnt from
such experiments, with a particular interest in the parity violating
polarized structure functions which have not yet been measured, but
might be experimentally accessible in a near future. We are thinking
of deep inelastic scattering processes at HERA or SLAC facilities, with
polarized electron, positron or muon beams and polarized nucleon targets.
Also the possibility of polarized proton beams appears promising
\ref\propol{CERN Courier {\bf 33} (1993) 6; 12}.
The extension of our results to neutrino initiated processes is trivial.
However, we do not consider these processes explicitely here because
of the technical difficulties in polarizing the large nucleon targets
needed for neutrino scatterings, which make these experiments
unrealistic at the moment.

There exist in the literature several works on the weak interaction
contribution to polarized deep inelastic scattering, with some
discrepancies among them
\nref\nash{C.~Nash, {\it Nucl. Phys.} {\bf B31} (1971)
419}\nref\der{E.~Derman, {\it Phys. Rev.} {\bf D7} (1973)
2755}\nref\kol{N.N.~Nikolaev, M.A.~Shifman and M.Z.~Shmatikov,
{\it JETP} {\bf 18} (1973) 39}\nref\ahm{M.A.~Ahmed and G.G.~Ross,
{\it Nucl. Phys.} {\bf B111}
(1976) 441}\nref\kaur{J.~Kaur, {\it Nucl. Phys.} {\bf B128} (1977)
219}\nref\jos{A.S.~Joshipura and P.~Roy, {\it Annals of Physics}
{\bf 104} (1977) 460}\nref\cah{R.N.~Cahn and F.J.~Gilman,
{\it Phys. Rev.} {\bf D17} (1978), 1313}\nref\bart{J.A.~Bartelski,
{\it Acta Phys. Pol.} {\bf B10} (1979)
10; 923}\nref\rep{N.S.~Craigie, K.~Hidaka, M.~Jacob and F.M.~Renard,
{\it Phys. Rep.} {\bf 99} (1983) 69}\nref\lampe{B.~Lampe, {\it Phys.
Lett.} {\bf B227} (1989) 469}\nref\vog{W.~Vogelsang and A.~Weber,
{\it Nucl. Phys.} {\bf B362} (1991) 3}\nref\jenk{E.~Jenkins, {\it Nucl.
Phys.} {\bf B354} (1991) 24}\nref\rav{V.~Ravishankar, {\it Nucl. Phys.}
{\bf B374} (1992) 309}\nref\gluon{P.~Mathews and V.~Ravindran,
{\it Int. Jour. Mod. Phys.} {\bf A7} (1992) 6371}\nref\ji{Xiangdong Ji,
{\it Nucl. Phys.} {\bf B402} (1993) 217}\nref\arg{D.E.~de
Florian \ea, La Plata {\it preprint} 93-03 (1993)}\refs{\nash - \arg};
our goal in this paper is that of presenting a complete and detailed
analysis of all contributions to charged lepton--nucleon interactions,
with explicit expressions of cross--sections and spin asymmetries,
with arbitrary spin directions, which
might be useful to experimentalists. Our emphasis is on the possible
experimental study of the nucleon structure in very high energy deep
inelastic scatterings within the Standard Model: we
suggest some measurements which, although difficult, might soon be
feasible and could allow to extract information on new polarized
structure functions. The partonic interpretation of all structure
functions is discussed and interesting combinations are indicated
which single out particular quark distributions; most of them
are independent of the axial anomaly, and allow an unambiguous
interpretation. A full study of the gluonic contributions
has been performed in Ref.\vog\ and a much more
formal approach to deep inelastic scattering with electroweak currents
can be found in Ref.\ji.

The paper is organized as follows. In Section 2 we introduce the
formalism and define the independent structure functions which appear
in the most general hadronic tensor describing the electro--weak
coupling of polarized nucleons to gauge bosons, both in case of neutral
and charged currents; we also give explicit formulas for
cross--sections with particular spin configurations. In Section 3 we
give the parton model expressions of the scaling structure functions
and in Section 4 we discuss how to perform useful measurements and to
combine particular cross--sections, in order to extract meaningful
combinations of quark and antiquark distribution functions.

\goodbreak
\vskip 18pt
\nobreak
\noindent
{\bf 2 - General formalism}
\vskip 12pt
Let us consider first the neutral current ($nc$) deep inelastic process
$\ell N \to \ell^\prime X$ with polarized charged leptons ($\ell$) and
nucleons ($N$). In lowest order perturbation theory the amplitude for
such a process is given by the sum of two contributions: the one photon
exchange ($M_\gamma$) and the $Z^0$ exchange ($\mz$). The differential
cross--section then reads:
\eqn\first{
{d^2 \sigma^{\ell N}_{nc} \over d\Omega dE^\prime} = {1\over 2\mn(4\pi)^2}
{E^\prime\over E} \times \vert M_\gamma + \mz \vert^2 \>,}
where $E$ and $E^\prime=E-\nu$ are, respectively, the initial and final
lepton energy in the laboratory frame where the nucleon of mass $\mn$
is at rest, and the scattered lepton, of mass $m_\ell$, emerges into solid
angle $d\Omega$. Terms proportional to $m_\ell/E$ or $m_\ell/E^\prime$ are
neglected throughout the paper.

Eq.\first\ receives contributions from a purely electromagnetic term
$(|M_\gamma|^2)$, a purely weak one $(|\mz|^2)$ and an interference
one ($M_\gamma M_{\spz}^* + \mz M_\gamma^*$), so that it can be cast in the
form
\eqn\crsec{
{d^3\sigma^{\ell N}_{nc} \over dx \, dy \, d\phi} = {y \alpha^2 \over 2Q^4}
\sum_{i=\gamma,\gamma \spz, \spz} L_{\mu\nu}^i W^{\mu\nu}_i \eta^i \>,}
where $Q^2 = -q^2$ is the squared four-momentum transfer,
$x = Q^2/(2\mn\nu)$, $y=\nu/E$ and $\phi$ is the azimuthal angle of the
final lepton. $L_{\mu\nu}^i$ is the usual leptonic tensor deriving from
the couplings of the $\gamma$ and the $Z^0$ to the lepton. For
negatively charged leptons ($e^-, \mu^-$) one has
\eqn\ltens{\eqalign{
L_{\mu\nu}^\gamma &= [\bar u(l^\prime) \gamma_\mu u(l)]^* \,
[\bar u(l^\prime) \gamma_\nu u(l)] \cr
&= 2[(l_\mu l^\prime_\nu + l^\prime_\mu l_\nu
- l \cdot l^\prime g_{\mu\nu}) - i\lambda \epsilon_{\mu\nu\alpha\beta}
l^\alpha l^{\prime \beta}] \cr
L_{\mu\nu}^{\gamma \spz} &= [\bar u(l^\prime) \gamma_\mu (\gv-\ga
\gamma_5) u(l)]^* \, [\bar u(l^\prime) \gamma_\nu u(l)] \cr
&= (\gv-\lambda \ga) L_{\mu\nu}^\gamma \cr
L_{\mu\nu}^{\spz} &= [\bar u(l^\prime) \gamma_\mu (\gv-\ga \gamma_5)u(l)]^*
\, [\bar u(l^\prime) \gamma_\nu (\gv-\ga \gamma_5)u(l)] \cr
&= (\gv-\lambda \ga)^2 L_{\mu\nu}^\gamma \cr}}
where $l$ and $\lambda = \pm1$ are, respectively, the four-momentum
and the helicity of the initial lepton and $l^\prime = l - q$ is the
four-momentum of the final lepton. For positively charged leptons
($e^+, \mu^+$) one should simply replace, in the above formula, $\ga$
with $-\ga$ \ref\lp{See, \eg, E.~Leader and E.~Predazzi,
{\it An Introduction to Gauge Theories and the `New Physics'},
Cambridge University Press (1985)}. In our notations
\eqn\gagv{\gv = -{1\over 2} + 2 \sin^2\theta_{\scriptscriptstyle W}
\quad\quad\quad
\ga=-{1\over 2}\,.}

The factors $\eta^i$ in Eq.\crsec\ collect some kinematical quantities,
coupling constants and the relative weights of different propagators,
namely:
\eqn\etas{\eqalign{
\eta^\gamma &= 1 \cr
\eta^{\gamma \spz} &= \left( {GM_{\spz}^2 \over 2\sqrt2\pi\alpha} \right)
\left( {Q^2 \over Q^2 + M_{\spz}^2} \right) \cr
\eta^{\spz} &= (\eta^{\gamma \spz})^2 \cr}}
where $G$ is the Fermi coupling constant and $\mz$ is the $Z^0$ mass.
Notice that $GM_{\spz}^2 / (2\sqrt2\pi\alpha) = (4\sin^2\tw
\cos^2\tw)^{-1} \simeq 4/3$.

Finally, the hadronic tensor $W^{\mu\nu}_i$ defines the coupling of the
electromagnetic ($J_\gamma$) and the weak ($J_{\spz}$) current to the nucleon
with its transition to all possible final states $X$ \lp,
\eqn\htensg{\eqalign{
W^{\mu\nu}_\gamma &= \sum_X <X|J^\mu_\gamma|N>^* <X|J^\nu_\gamma|N>
(2\pi)^3 \delta (\px-p-q) \cr
W^{\mu\nu}_{\gamma \spz} &= \sum_X \bigl[ <X|J^\mu_{\spz}|N>^*
<X|J^\nu_\gamma|N> \cr
& \quad\quad + <X|J^\mu_\gamma|N>^* <X|J^\nu_{\spz}|N>
\bigr] (2\pi)^3 \delta (\px-p-q) \cr
W^{\mu\nu}_{\spz} &= \sum_X <X|J^\mu_{\spz}|N>^* <X|J^\nu_{\spz}|N>
(2\pi)^3 \delta (\px-p-q) \cr}}
where $\px$ is the total four-momentum of the state $X$. Exploiting
Lorentz and CP invariance the hadronic tensor can be expressed
in terms of $q^\mu$, the nucleon four-momentum $p^\mu$, its spin $S^\mu$
and 8 independent structure functions \bart, as

\eqn\htens{\eqalign{
{1\over 2\mn} W^i_{\mu\nu} &= -{g_{\mu\nu} \over \mn} \, F_1^i
+{p_\mu p_\nu\over \mn (p\cdot q)}\,F_2^i \cr
& +i{\epsilon_{\mu\nu\alpha\beta}\over 2(p\cdot q)} \left[
{p^\alpha q^\beta \over \mn} \, F_3^i + 2 q^\alpha S^\beta \, g_1^i -
4x p^\alpha S^\beta \, g_2^i \right] \cr
&-{p_\mu S_\nu + S_\mu p_\nu \over 2(p\cdot q)} \, g_3^i +
{S\cdot q \over (p\cdot q)^2} \, p_\mu p_\nu \, g_4^i
+{S\cdot q\over p\cdot q} \, g_{\mu\nu} \, g_5^i \,.\cr}}
Notice that terms proportional to $q^\mu$ or $q^\nu$ can be dropped in
the definition of $W^i_{\mu\nu}$ because they give no contribution
(in the $m_\ell/E \to 0$ limit) when contracted with $L_i^{\mu\nu}$.
Different definitions of the hadronic tensor appearing in the literature
differ from ours due to these terms. In particular the coefficient of
$g_2$ could be written in a more familiar way using the identity
\ref\heim{R.~L.~Heimann,
{\it Nucl. Phys.} {\bf B64} (1973) 429}
\eqn\iden{\eqalign{
\epsilon_{\mu\nu\alpha\beta} \, p^\alpha S^\beta &=
{\epsilon_{\mu\nu\alpha\beta} \over 2x(p\cdot q)}
\left[ (q\cdot S) q^\alpha p^\beta - (p\cdot q) q^\alpha S^\beta
\right] \cr
&- \left( q_\mu \epsilon_{\nu\alpha\beta\gamma}
- q_\nu \epsilon_{\mu\alpha\beta\gamma} \right)
p^\alpha q^\beta S^\gamma \cr}}
and dropping the last two terms.

The structure functions $F_j(Q^2, p\cdot q)$ and $g_j(Q^2, p\cdot q)$
are expected to scale in the large $Q^2$ limit and to depend only
on the Bjorken scaling variable $x = Q^2/ 2(p\cdot q)$.
The $F_j$ are the unpolarized structure functions and the $g_j$ the
polarized ones: when averaging over the nucleon spin one computes
$W_{\mu\nu}(p,q,S) + W_{\mu\nu}(p,q,-S)$ and all terms proportional
to $g_j$ cancel out. In Eq.\htens\ we have allowed for parity
violations and indeed $W_{\mu\nu}^i$ is a mixture of second rank
tensors and pseudotensors; in case of pure electromagnetic interactions
($i=\gamma$) parity is conserved and one has
\eqn\fgelm{
F_3^\gamma = g_3^\gamma = g_4^\gamma = g_5^\gamma = 0.}
$F_3$, $g_3$, $g_4$ and $g_5$ only contribute to parity violating
interactions and are often referred to as the parity violating
structure functions.

The structure functions defined according to Eq.\htens\ are
such that one recovers the usual ones \lp\ when considering
unpolarized DIS, and electromagnetic polarized DIS.

We shall now consider some particular cases of Eq.\crsec, corresponding
to specific nucleon spin configurations. Let us start from
a longitudinally polarized nucleon. If we choose the $z$-axis as the
direction of motion of the incoming lepton we then have
\eqn\long{\eqalign{
l^\mu &= E(1,0,0,1) \quad\quad l^{\prime \mu} =
E^\prime(1,\sin\theta\cos\phi,\sin\theta\sin\phi,\cos\theta) \cr
p^\mu &= (\mn,0,0,0) \quad\>\> S^\mu=S_L^\mu=(0,0,0,1) \>, \cr}}
and we obtain (after integration over the azimuthal angle $\phi$):
\eqn\cslong{\eqalign{
{d^2\sigma^{\ell N}_{nc} \over dx \, dy}&(\lambda, S=S_L)=
4 \pi \mn Ey \, {\alpha^2\over Q^4} \sum_i \eta^i C^i \cr
&\times \biggl\{ 2xy\,F_1^i +
{2\over y} \left( 1-y-{xy\mn\over 2E}\right) (F_2^i + g_3^i) \cr
&-2\lambda x\left( 1-{y\over2} \right)F_3^i
-2\lambda x\left(2-y-{xy\mn\over E}\right)g_1^i
+4\lambda {x^2\mn\over E}\, g_2^i \cr
&-{2\over y}\left( 1+{x\mn\over E}\right)\left(1-y-{xy\mn\over 2E}
\right)g_4^i +2xy\left(1+{x\mn\over E}\right) g_5^i \biggl\}\cr}}
where, for negatively charged leptons,
\eqn\Ci{
C^\gamma=1 \quad\quad C^{\gamma \spz}=(\gv-\lambda \ga)
\quad\quad C^{\spz}=(\gv-\lambda \ga)^2\,,}
and for positively charged leptons one simply replaces $\ga$ with
$-\ga$; $\gv$ and $\ga$ are as in Eq.\gagv.
Notice that when the lepton flips its helicity $\lambda$ changes sign
and when the nucleon flips its spin all terms containing a polarized
structure funtion $g_j \> (j=1,2,..,5)$ also change sign. Upon averaging
over $\lambda$ and $S$ one obtains the unpolarized cross--section
\eqn\csunp{\eqalign{
{d^2\sigma^{\ell N}_{nc} \over dx \, dy}({\rm unp.}) &=
{1\over 4} \sum_{\lambda, S}
{d^2\sigma^{\ell N}_{nc} \over dx \, dy}(\lambda, S) \cr
&= 4 \pi \mn Ey \, {\alpha^2\over Q^4} \sum_i \eta^i C^i
\left\{ 2xy \, F_1^i +{2\over y} \left(1-y-{xy\mn\over 2E}\right)
F_2^i\right\}\,.\cr}}

In the case of nucleons with transverse polarization, \ie\ with a spin
orthogonal to the lepton direction ($z$-axis) at an angle $\alpha$ to
the $x$-axis,
we have
\eqn\tran{
S^\mu = S^\mu_T = (0, \cos\alpha, \sin\alpha, 0) \,}
which, using Eqs.\crsec, \ltens\ and \htens, yields
\eqn\cstran{\eqalign{
{d^3\sigma^{\ell N}_{nc} \over dx \, dy \, d\phi}&(\lambda, S=S_T)=
2\mn Ey \, {\alpha^2\over Q^4} \sum_i \eta^i C^i \cr
&\times \biggl\{ 2xy\,F_1^i +
{2\over y} \left( 1-y-{xy\mn\over 2E}\right)F_2^i
- 2\lambda x\left( 1-{y\over2} \right)F_3^i \cr
&+{\sqrt{xy\mn[2(1-y)E-xy\mn]}\over E} \, \cos(\alpha-\phi) \cr
&\times \Bigl[ -2\lambda xg_1^i -4\lambda {x\over y} \, g_2^i
+{1\over y} \, g_3^i
+{2\over y^2} \left( 1-y-{xy\mn\over 2E}\right) g_4^i
-2xg_5^i \Bigr] \biggl\}\,.\cr}}
Again, a flip in the lepton helicity results in a change of the sign
of $\lambda$ (including in the expressions for $C^i$, Eq.\Ci) and
a flip of the nucleon spin induces a change of sign in all terms
containing a polarized structure function $g_j$.

Before turning to a discussion of the parton model interpretation of
the structure functions we consider also the case of charged current
($cc$) deep inelastic processes $\ell^\mp N \to \nu (\bar \nu) \> X$,
which proceed, at lowest perturbative order, through the exchange of a
charged vector boson $W^\mp$. This case resembles the $Z^0$
contribution to the neutral current process, with the assignement
$\gv=\ga=1$. In fact Eq.\first\ now reads
\eqn\firstcc{
{d^2 \sigma^{\ell N}_{cc} \over d\Omega dE^\prime} = {1\over 2\mn(4\pi)^2}
{E^\prime\over E} \times \vert \mw \vert^2 \>,}
and Eq.\crsec\ changes into
\eqn\crseccc{
{d^3\sigma^{\ell N}_{cc} \over dx \, dy \, d\phi} = {y \alpha^2 \over 2Q^4}
L_{\mu\nu}^{\spw} W^{\mu\nu}_{\spw} \eta^{\spw} \>,}
where, for a negatively charged lepton $\ell^-$ (which couples to a
$W^-$),
\eqn\ltenscc{
L_{\mu\nu}^{\spw^-} = [\bar u(l^\prime) \gamma_\mu (1-\gamma_5)u(l)]^*
\, [\bar u(l^\prime) \gamma_\nu (1-\gamma_5)u(l)]
= (1-\lambda)^2 L_{\mu\nu}^\gamma \,.}
Eq.\ltenscc\ shows that fast $\ell^-$ leptons only couple to a $W$
if they have a negative helicity ($\lambda=-1$).
Eq.\crseccc\ is completed by
\eqn\etacc{
\eta^{\spw} = {1\over 2}\left( {GM_{\spw}^2 \over 4\pi\alpha}
{Q^2 \over Q^2 + M_{\spw}^2} \right)^2\,,}
where $\mw$ is the vector boson mass, and
$W_{\mu\nu}^{\spw} = W_{\mu\nu}^i$, Eq.\htens. For a positively charged
lepton ($\ell^+$) one simply changes, as usual, the sign of the axial
coupling $\gamma^\mu\gamma_5$, that is one replaces $\lambda$ with
$-\lambda$ in Eq.\ltenscc\ to get $L_{\mu\nu}^{\spw^+}=
(1+\lambda)^2 L_{\mu\nu}^\gamma$.

Eqs.\cslong\ and \cstran\ then hold also for the charged current
interaction, with no $\sum_i$, the factor $\eta$ as in Eq.\etacc\
and $C=(1-\lambda)^2$ for $\ell^-$, $C=(1+\lambda)^2$ for $\ell^+$;
they will be used in Section 4 when discussing meaningful possible
measurements and we will drop the index $nc$. All of the previous
formulae can also be easily modified to describe neutrino initiated
processes, which we do not consider here.

\goodbreak
\vskip 18pt
\nobreak
\noindent
{\bf 3 - Structure functions in the naive quark-parton model}
\vskip 12pt
According to the naive quark-parton model the lepton interaction with
the nucleon is just the incoherent sum of all elementary
interactions of the gauge bosons $\gamma, W^\pm$ and $Z^0$
with the quarks, supposed to be {\it free} constituents,
each of which carries a fraction $x$ of the proton four-momentum,
$k^\mu = xp^\mu$. This leads to \lp:
\eqn\htensp{
W^{i,\mu\nu} = \sum_q {1\over 2x\mn\nu} \, \Bigl[
\omega_{q,S}^{i,\mu\nu} \, q^P + \omega_{q,-S}^{i,\mu\nu} \, q^A +
\omega_{\bar q,S}^{i,\mu\nu} \, \bar q^P + \omega_{\bar q,-S}^{i,\mu\nu}
\, \bar q^A \Bigr]}
where $q^{P(A)}(x)$ is the number density of quarks with flavour $q$
and spin parallel (antiparallel) to the nucleon spin; similarly for
$\bar q^{P,A}(x)$, which refer to antiquarks.
$\omega_{q,s}^{i,\mu\nu}$ is the quark tensor, analogous to the
leptonic tensor $L^{\mu\nu}$, and the index $i$ refers, as usual, to
the different interaction contributions, $i= \gamma, \gamma Z, Z, W$:
\eqn\qtens{\eqalign{
\omega_{q,s}^{\gamma,\mu\nu} &= \sum_{s^\prime} e_q^2 \,
[\bar u(k^\prime,s^\prime) \gamma^\mu u(k,s)]^* \,
[\bar u(k^\prime,s^\prime) \gamma^\nu u(k,s)] \cr
\omega_{q,s}^{\gamma \spz,\mu\nu} &= \sum_{s^\prime} e_q \,
[\bar u(k^\prime,s^\prime) \gamma^\mu (\gv-\ga\gamma_5)_q u(k,s)]^* \,
[\bar u(k^\prime,s^\prime) \gamma^\nu u(k,s)] \cr
&+ \sum_{s^\prime} e_q \,
[\bar u(k^\prime,s^\prime) \gamma^\mu u(k,s)]^* \,
[\bar u(k^\prime,s^\prime) \gamma^\nu (\gv-\ga\gamma_5)_q u(k,s)] \cr
\omega_{q,s}^{\spz,\mu\nu} &= \sum_{s^\prime}
[\bar u(k^\prime,s^\prime) \gamma^\mu (\gv-\ga\gamma_5)_q u(k,s)]^* \,
[\bar u(k^\prime,s^\prime) \gamma^\nu (\gv-\ga\gamma_5)_q u(k,s)] \cr
\omega_{q,s}^{\spw,\mu\nu} &= \sum_{s^\prime, q^\prime}
[\bar u(k^\prime,s^\prime) \gamma^\mu (1-\gamma_5) u(k,s)]^* \,
[\bar u(k^\prime,s^\prime) \gamma^\nu (1-\gamma_5) u(k,s)] \,
|({\rm K.M.})_{qq^\prime}|^2 \cr}}
where $k$, $k^\prime = k+q$, $s$ and $s^\prime$ are respectively the
momentum and spin four-vectors of the initial and final quarks;
the initial quark is not necessarily supposed to be in a helicity state
but its spin is taken to be either parallel to the nucleon spin ($s=S$)
or antiparallel ($s=-S$). $(\gv)_q$ and $(\ga)_q$ are the vector and
axial couplings of the quark of flavour $q$ to the $Z^0$ and $e_q$ is
the quark charge in units of the proton charge.
In case of charged current, negative charge leptons only
couple to $u$-type quarks (or $\bar d$-type antiquarks) and positive
charge leptons couple to $d$-type quarks (or $\bar u$-type antiquarks);
one has also to take into account the proper Cabibbo-Kobayashi-Maskawa
matrix elements occurring in the transition coupling from a flavour
$q$ to a flavour $q^\prime$. However, if we consider the contribution
of four flavours ($u$, $d$, $s$ and $c$), one always has
$\sum_{q^\prime} |({\rm K.M.})_{qq^\prime}|^2 = \cos^2\theta_c +
\sin^2\theta_c = 1$, where $\theta_c$ is the Cabibbo angle.

Explicit expressions of Eqs.\qtens\ can be obtained from the general
form
\eqn\qtensgen{\eqalign{
\omega_{q,s}^{\mu\nu} &= \sum_{s^\prime}
[\bar u(k^\prime,s^\prime) \gamma^\mu (v_1 - a_1\gamma_5) u(k,s)]^* \,
[\bar u(k^\prime,s^\prime) \gamma^\nu (v_2 - a_2\gamma_5) u(k,s)] \cr
&=2\,(a_1a_2+v_1v_2) \left[ 2k^\mu k^\nu-k\cdot q \>g^{\mu\nu}\right]
-4a_1a_2 \, m_q^2 \, g^{\mu\nu} \cr
&-2v_1a_2 \, m_q \left[
2 k^\mu s^\nu -s\cdot q \> g^{\mu\nu} \right]
-2a_1v_2 \, m_q \left[
2 s^\mu k^\nu -s\cdot q \> g^{\mu\nu} \right] \cr
&+2i\epsilon^{\mu\nu\alpha\beta} \left[ (v_1a_2+a_1v_2) \,
k_\alpha q_\beta + 2a_1a_2 \, m_q \, k_\alpha s_\beta +
(a_1a_2+v_1v_2) m_q \, q_\alpha s_\beta
\right] \cr}}
by properly fixing the values of $v_{1,2}$ and $a_{1,2}$;
$m_q$ is the quark mass and we have dropped terms
proportional to $q^\mu$ or $q^\nu$ which, when contracted with the
leptonic tensor $L_{\mu\nu}$, give contributions
proportional to $m_\ell/E$, consistently neglected in the whole paper.
If the quark has opposite spin it suffices to change the sign of $s$.

The antiquark tensors $\omega_{\bar q,s}^{i,\mu\nu}$ are defined
exactly as in Eq.\qtens\ with the only replacement $\gamma_5 \to
-\gamma_5$; Eq.\qtensgen\ can then be exploited also for antiquarks and
gives $\omega_{\bar q,s}^{\mu\nu}$ as a result of the usual replacement
in the axial couplings, $a_{1,2} \to - a_{1,2}$.

Eqs.\htensp-\qtensgen\ give the quark-parton model predictions
for the hadronic tensor $W^{i,\mu\nu}$; by comparing them with the
general expression, Eq.\htens, one obtains the quark-parton model
results for the nucleon structure functions. For completeness we list
all of them here, starting from the electromagnetic case ($i=\gamma$):
\eqn\fgelm{\eqalign{
F_1^\gamma &= {1\over 2} \sum_q e_q^2 (q + \bar q)
\quad\quad F_2^\gamma = 2x F_1^\gamma \cr
g_1^\gamma &= {1\over 2} \sum_q e_q^2 (\Delta q + \Delta \bar q)
\quad\quad g_2^\gamma = 0 \cr}}
where $q = q^P + q^A$ is the number density of quarks of flavour $q$
and $\Delta q = q^P - q^A$; analogously for antiquarks.

The interference contribution ($i=\gamma Z$) is:
\eqn\fgint{\eqalign{
F_1^{\gamma \spz} &= \sum_q e_q (\gv)_q (q + \bar q)
\quad\quad F_2^{\gamma \spz} = 2xF_1^{\gamma \spz} \cr
F_3^{\gamma \spz} &= 2 \sum_q e_q (\ga)_q (q - \bar q) \cr
g_1^{\gamma \spz} &= \sum_q e_q (\gv)_q (\Delta q +
\Delta \bar q) \cr
g_2^{\gamma \spz} &= g_4^{\gamma \spz} = 0 \cr
g_3^{\gamma \spz} &= 2x \sum_q e_q (\ga)_q (\Delta q - \Delta \bar q)
\quad\quad  2 x g_5^{\gamma \spz} = g_3^{\gamma \spz} \cr}}
and the purely weak interaction ($i=Z$) leads to:
\eqn\fgweak{\eqalign{
F_1^{\spz} &= {1\over 2} \sum_q (g_{\spv}^2 + g_{\spa}^2)_q (q + \bar q)
\quad\quad F_2^{\spz} = 2 x F_1^{\spz} \cr
F_3^{\spz} &= 2 \sum_q (\gv \ga)_q (q - \bar q) \cr
g_1^{\spz} &= {1\over 2} \sum_q (g_{\spv}^2 + g_{\spa}^2)_q (\Delta q+
\Delta \bar q)\cr
g_2^{\spz} &= -{1\over 2}\sum_q (g_{\spa}^2)_q (\Delta q + \Delta \bar q)
\cr
g_3^{\spz} &= 2 x \sum_q (\gv \ga)_q (\Delta q - \Delta \bar q)
\quad\quad g_4^{\spz} = 0 \quad\quad 2 x g_5^{\spz} = g_3^{\spz} \cr}}

In case of charged current ($i=W$), on performing explicitely the
$\sum_q$, one obtains, for $\ell^-N \to \nu X$ processes:
\eqn\fgcc{\eqalign{
F_1^{\spw^-} &= u + c + \bar d + \bar s
\quad\quad F_2^{\spw^-} = 2 x F_1^{\spw^-} \cr
F_3^{\spw^-} &= 2(u + c - \bar d - \bar s) \cr
g_1^{\spw^-} &= (\Delta u + \Delta c + \Delta \bar d + \Delta \bar s)
\quad\quad 2g_2^{\spw^-} = -g_1^{\spw^-} \cr
g_3^{\spw^-} &= 2 x (\Delta\ u + \Delta c - \Delta \bar d -
\Delta \bar s) \quad\quad g_4^{\spw^-} = 0
\quad\quad 2 x g_5^{\spw^-} = g_3^{\spw^-} \cr}}
where $u$ stays for the number density of quarks $u$ and so on.
$\ell^+N \to \bar \nu X$ processes probe different quark flavours
and one obtains the corresponding expressions of the structure
functions $F_j^{\spw^+}$ and $g_j^{\spw^+}$ by the flavour interchanges
$d \leftrightarrow u$ and $s \leftrightarrow c$ in the above Eq. \fgcc.

Notice that in the quark--parton model the structure functions $g^i_4$
are always zero and one finds $g_3^i=2xg_5^i$ for any $i=\gamma,
\gamma Z, Z, W$. Indeed in the Bjorken limit a Callan--Gross--like
relation holds, independently of the actual model of the nucleon
employed in the calculation and of the kind of local interaction,
provided that the elementary  point--like constituents are fermions,
as follows from an analysis of the helicity amplitudes
\der\ or from the OPE of the hadronic
tensor \rav: $g_3-g_4=2x g_5$. The quark--parton model only sets $g_4=0$.
The function $g_2^i$ are nonzero only for pure weak
interactions (both $nc$ and $cc$ ones). It is interesting to note that,
in case of neutral current, the integral of $g_2^{\spz}$ appears to be
directly proportional to the the total spin carried by the quarks and
antiquarks, as can be seen from Eqs.\fgweak\ and the fact that
$g_{\spa}^2 = 1/4$ for any quark flavour. Its experimental determination,
however, appears to be very difficult, as can be seen from Eqs.\cslong\
and \cstran\ and will be discussed in the next Section.

This completes our summary of the naive quark-parton model predictions
for the different structure functions; we should now tackle the problem
of experimental measurements which could single out the various
contributions and possibly give new information on the polarized quark
distributions inside the nucleons.

\goodbreak
\vskip 18pt
\nobreak
{\bf 4 - Experimental measurements}
\noindent
\vskip 12pt
We would like to suggest some ideal experimental measurements which could
provide additional precious information on the quark and spin content
of nucleons. We assume that high energy longitudinally polarized
electron, positron or muon beams are available, as well as polarized
proton and neutron targets or proton beams, and consider some suitable
combinations of
cross--sections. Such measurements might become affordable in the near
future at HERA, CERN or SLAC facilities.

Let us define the nucleon spin asymmetries:
\eqn\lsasdef{
\Delta^L\sigma^{\ell N}(\lambda)
\equiv {d^2\sigma^{\ell N} \over dx \, dy} (\lambda, S=S_L)
     - {d^2\sigma^{\ell N} \over dx \, dy} (\lambda, S=-S_L)}
and
\eqn\tsasdef{
\Delta^T\sigma^{\ell N}(\lambda)
\equiv {d^3\sigma^{\ell N} \over dx \, dy \, d\phi} (\lambda, S=S_T)
     - {d^3\sigma^{\ell N} \over dx \, dy \, d\phi} (\lambda, S=-S_T)\,}
which single out the polarized structure functions.
Eqs. \cslong\ and \cstran, together with the parton model results
$g_4^i = 0$ and $g_3^i = 2xg_5^i$, yield
\eqn\lsas{\eqalign{
\Delta^L\sigma^{\ell N}(\lambda) &= 16\pi \mn E \, {\alpha^2\over Q^4}
\sum_i \eta^i C^i \cr
& \times \Bigl\{ -\lambda xy \left( 2-y-{xy\mn\over E} \right) g_1^i
+2\lambda {x^2y\mn\over E}\, g_2^i \cr
&+ x \left[ y^2 + (1-y) \left( 2 - {xy\mn\over E} \right) \right]
g_5^i \Bigr\}\cr}}
and
\eqn\tsas{\eqalign{
\Delta^T\sigma^{\ell N}(\lambda) &= 8\mn \,{\alpha^2\over Q^4} \,
\cos(\alpha-\phi) \sqrt{xy\mn[2(1-y)E-xy\mn]} \cr
&\times \sum_i \eta^i C^i \Bigl\{
-\lambda xy \, g_1^i - 2 \lambda x\, g_2^i
+ x(1-y) \, g_5^i \Bigl\} \cr}}
which hold, depending on the values of the factors $\eta^iC^i$, both
for neutral and charged current processes. For the $nc$ case the
$\eta^i$ are given in Eq.\etas\ and the $C^i$ in Eq.\Ci; for the $cc$
case there is only one term in the sum, with $\eta$ given in Eq.\etacc\
and $C=(1\mp \lambda)^2$ respectively for $\ell^\mp$. Notice that the
transverse asymmetry \tsas\ is suppressed by a factor $\sqrt{\mn/E}$
with respect to the longitudinal one, Eq.\lsas, and the unpolarized
cross--section, Eq.\csunp; this might make its measurement problematic.

Let us now discuss the charged current case. From
 Eqs.\lsas, \tsas\ and the parton model results \fgcc\ one
has, for negatively charged leptons ($\ell^-$) with helicity $\lambda=-1$
or for positively charged ones ($\ell^+$) with helicity $\lambda=1$:
\eqn\sascc{\eqalign{
\Delta^L\sigma^{\ell^\mp N}_{cc} &= 64\pi \mn E \, {\alpha^2 \over Q^4}
\, \eta^{\spw} \times \Bigl\{ \pm xy
\left[ 2-y+{x\mn\over E}(1-y)\right] g_1^{\spw^\mp} \cr
&+ x \left[ y^2 + (1-y) \left( 2 - {xy\mn\over E} \right) \right]
g_5^{\spw^\mp} \Bigr\} , \cr
\Delta^T\sigma^{\ell^\mp N}_{cc} &= 32 \mn {\alpha^2 \, \over Q^4} \,
\eta^{\spw} \sqrt{xy\mn[2(1-y)E-xy\mn]} \cos(\alpha-\phi) \cr
&\times x(1-y) \left( \mp g_1^{\spw^\mp} + g_5^{\spw^\mp} \right). \cr}}

Eqs.\sascc\ show how separate measurements of
$\Delta^L\sigma^{\ell^\mp}_{cc}$ and $\Delta^T\sigma^{\ell^\mp}_{cc}$
could allow a determination of $g_1^{\spw^\mp}$ and $g_5^{\spw^\mp}$. By
suitable combinations of these structure functions one could extract
meaningful information on the spin content of the nucleon. For
example:
\eqn\gpgcc{\eqalign{
g_1^{\spw^-} + g_1^{\spw^+}  &= (\Delta u + \Delta\bar u + \Delta d +
\Delta\bar d + \Delta s + \Delta\bar s + \Delta c +\Delta \bar c) \cr
&= \sum_q (\Delta q + \Delta\bar q) \cr}}
and
\eqn\gmgcc{
g_5^{\spw^-} - g_5^{\spw^+} = (\Delta u + \Delta\bar u + \Delta c +
\Delta \bar c) - (\Delta d + \Delta\bar d + \Delta s + \Delta\bar s)\,.}
Notice that the combination \gpgcc\ supplies another way of getting
information on the total spin carried by quarks and antiquarks;
moreover, by scattering on an isoscalar target one could obtain
information on the strange quark polarization,
\eqn\gmgccis{
(g_5^{\spw^-} - g_5^{\spw^+})_p + (g_5^{\spw^-} - g_5^{\spw^+})_n =
2(\Delta c + \Delta\bar c - \Delta s - \Delta \bar s)_p}
where the indexes $p$ and $n$ stay respectively for proton and neutron.

Let us now consider neutral current interactions. To simplify our
results we notice that in the high energy region one can keep only
leading terms in  $\mn/E$; moreover, from Eq.\etas\ and for $Q^2$
values up to $\sim 10^3$ GeV$^2$ one has
\eqn\etasin{
\eta^{\spz} \ll \eta^{\gamma \spz} \ll \eta^\gamma \,.}
Eqs.\lsas\ and \tsas\ and the parton model results \fgelm-\fgweak\
then give, with a very good approximation:
\eqn\sasnc{\eqalign{
\Delta^L \sigma^{\ell N}_{nc}(\lambda=1) &=
-16\pi \mn E \, {\alpha^2\over Q^4} \, xy(2-y) \, g_1^\gamma \cr
\Delta^T \sigma^{\ell N}_{nc}(\lambda=1) &=
-8 \mn \, {\alpha^2\over Q^4} \, \cos(\alpha-\phi) \sqrt{2xy\mn E(1-y)}
\, xy \, g_1^\gamma \cr}}
which allow a direct measurement of $g_1^\gamma$ and show how, in our
kinematical region, such asymmetries are still dominated by
electromagnetic interactions.

In order to extract information on the other polarized structure
functions, or simple combinations of few of them, in neutral current
processes, it is convenient to introduce further quantities:
\eqn\defds{\eqalign{
\Sigma_L(\lambda) \equiv \Delta^L\sigma^{\ell^-N}_{nc}(\lambda)
                       + \Delta^L\sigma^{\ell^+N}_{nc}(\lambda) \cr
\Sigma_T(\lambda) \equiv \Delta^T\sigma^{\ell^-N}_{nc}(\lambda)
                       + \Delta^T\sigma^{\ell^+N}_{nc}(\lambda) \cr
D_L(\lambda) \equiv \Delta^L\sigma^{\ell^-N}_{nc}(\lambda)
                  - \Delta^L\sigma^{\ell^+N}_{nc}(\lambda) \cr
D_T(\lambda) \equiv \Delta^T\sigma^{\ell^-N}_{nc}(\lambda)
                  - \Delta^T\sigma^{\ell^+N}_{nc}(\lambda) \cr}}
whose full explicit expressions can be found from Eqs.\lsas\ and
\tsas. At large energy and in the $Q^2$ range of validity of
Eq.\etasin\ they can be combined, using the parton model results,
and noticing, Eq.\gagv, that for the lepton $\gv \simeq -0.04$
whereas $\ga = -0.5$, to give:
\eqn\comb{\eqalign{
\Sigma_L(\lambda=1) + \Sigma_L(\lambda=-1) &=
64\pi \mn E \, {\alpha^2\over Q^4} \, x(2-2y+y^2) \left\{ \gv
\eta^{\gamma \spz} \, g_5^{\gamma \spz} + g_{\spa}^2 \eta^{\spz} \,
g_5^{\spz} \right\} \cr
\Sigma_T(\lambda=1) + \Sigma_T(\lambda=-1) &=
32 \mn \, {\alpha^2\over Q^4} \, \cos(\alpha-\phi) \sqrt{2xy\mn E(1-y)}
\, x(1-y) \cr
&\times \left\{ \gv \eta^{\gamma \spz} \, g_5^{\gamma \spz}
+ g_{\spa}^2 \eta^{\spz} \, g_5^{\spz} \right\} \cr
D_L(\lambda=1) - D_L(\lambda=-1) &= -64\pi \mn E \, {\alpha^2\over Q^4}
\, x(2-2y+y^2) \, \ga \eta^{\gamma \spz} \, g_5^{\gamma \spz} \cr
D_L(\lambda=1) + D_L(\lambda=-1) &= 64\pi \mn E \, {\alpha^2\over Q^4}
\, xy(2-y) \, \ga \eta^{\gamma \spz} \, g_1^{\gamma \spz} \cr
D_T(\lambda=1) - D_T(\lambda=-1) &= -32 \mn \, {\alpha^2\over Q^4} \,
\cos(\alpha-\phi) \sqrt{2xy\mn E(1-y)} \cr
&\times x(1-y)\, \ga \eta^{\gamma \spz} \, g_5^{\gamma \spz} \cr
D_T(\lambda=1) + D_T(\lambda=-1) &= 32 \mn \, {\alpha^2\over Q^4} \,
\cos(\alpha-\phi) \sqrt{2xy\mn E(1-y)} \,
xy \, \ga \eta^{\gamma \spz} \, g_1^{\gamma \spz} \cr}}

 \ From a measurement of these quantities and the spin asymmetries
\sasnc\ one can obtain information on the structure functions
$g_1^\gamma$, $g_1^{\gamma \spz}$, $g_5^{\gamma \spz}$ and $g_5^{\spz}$.
Their parton model expressions, using
Eqs.\fgelm-\fgweak\ and the quark weak coupling constants,
\eqn\qcs{\eqalign{
(\gv)_{u,c} &= {1\over 2} - {4\over 3} \sin^2\tw
\quad\quad\quad (\ga)_{u,c} = {1\over 2} \cr
(\gv)_{d,s} &= -{1\over 2} + {2\over 3} \sin^2\tw
\quad\quad (\ga)_{d,s} = -{1\over 2} \,,\cr}}
are given by
\eqn\finalq{\eqalign{
g_1^\gamma &= {2\over 9} (\Delta u + \Delta c + \Delta \bar u
+ \Delta \bar c) + {1\over 18} (\Delta d + \Delta s + \Delta \bar d
+ \Delta \bar s) \cr
g_1^{\gamma \spz} &= \left( {1\over 3} - {8\over 9} \sin^2\tw \right)
(\Delta u + \Delta c + \Delta \bar u + \Delta \bar c) \cr
&+ \left( {1\over 6} - {2\over 9} \sin^2\tw \right)
(\Delta d + \Delta s + \Delta \bar d + \Delta \bar s)
\simeq {1\over 9} \sum_q (\Delta_q + \Delta_{\bar q}) \cr
g_5^{\gamma \spz} &= {1\over 6} \left[
2\,(\Delta u + \Delta c - \Delta \bar u - \Delta \bar c) +
(\Delta d + \Delta s - \Delta \bar d - \Delta \bar s) \right] \cr
g_5^{\spz} &= {1\over 2} \left( {1\over 2}-{4\over 3} \sin^2\tw \right)
(\Delta u + \Delta c - \Delta \bar u - \Delta \bar c) \cr
&+ {1\over 2} \left( {1\over 2} - {2\over 3} \sin^2\tw \right)
(\Delta d + \Delta s - \Delta \bar d - \Delta \bar s) \cr}}
where the approximate equality holds when assuming $\sin^2\tw
=1/4$.

In principle, a measurement of the above four structure functions allows
to extract information on the four combinations of quark distribution
functions $\Delta u + \Delta c$, $\Delta d + \Delta s$, $\Delta \bar u
+ \Delta \bar c$ and $\Delta \bar d + \Delta \bar s$.

Throughout the paper we have assumed the initial leptons to be in pure
helicity states (either $\lambda=1$ or $\lambda=-1$); however, our
formulae can easily be modified to take into account leptons with an
arbitrary average helicity
\eqn\avhel{
\langle\ \lambda\ \rangle = P(1) - P(-1),}
where $P(\lambda)$ is the probability of helicity $\lambda$. By
suitably tuning the value of $\langle\lambda\rangle$ one could extract single
structure functions. For example, from
\eqn\avhelcs{
\Delta^L\sigma^{\ell N}(\langle\lambda\rangle) =
P(1) \, \Delta^L\sigma^{\ell N}(\lambda=1) +
P(-1) \, \Delta^L\sigma^{\ell N}(\lambda=-1)\,}
one has
\eqn\spavhel{\eqalign{
&\Delta^L\sigma^{\ell^-N}_{nc}(\langle\lambda\rangle={\gv\over \ga})
+ \Delta^L\sigma^{\ell^+N}_{nc}(\langle\lambda\rangle=-{\gv\over \ga})= \cr
&=32\pi \mn E {\alpha^2\over Q^4}x ( 2-2y+y^2)
(g_{\spa}^2 - g_{\spv}^2)\, \eta^{\spz} \, g_5^{\spz}. \cr}}

Information on $g_1^{\gamma \spz}$ and $g_5^{\gamma \spz}$ could also be
obtained, as pointed out in \ref\bil{S.M. Bilenky, N.A. Dadajan and E.H.
Hristova, {\it Sov. J. Nucl. Phys.} {\bf 21} (1975) 657},
 by scattering unpolarized
leptons off longitudinally polarized nucleons; in this case indeed
there is no spin dependent e.m. contribution, and from Eqs.(29) and
(35) one has:
\eqn\pippo{\eqalign{
 \Delta^L \sigma^{\ell^- N}_{nc} (\langle \lambda\rangle=0) =
16 \pi \mn E {\alpha^2 \over Q^4} \eta^{\gamma \spz}\ x
\left\{ y(2-y) \ga \ g_1^{\gamma \spz} +
(2-2y+y^2) \gv \ g_5^{\gamma \spz} \right\}. } }

Eqs.\sascc, \sasnc, \comb, \spavhel\ and \pippo\ are only few out of the many
particular combinations of cross--sections which one might think of
measuring in order to extract new information on the spin structure
of nucleons; many others could be derived from our formulae and have
not been explicitly presented here. Also, we have written our results
always assuming the validity of the parton model predictions for the
structure functions, like $g_4^i=0$, $g_3^i=2xg_5^i$, $g_2^\gamma=0$,
{\it etc.}. Of course, from Eqs.\cslong\ and \cstran, one could obtain
the most general expressions for our combination of cross--sections,
independently of the parton model results. Such general expressions are
somewhat more complicated than the ones shown here in that they contain
more structure functions.

All of the measurements suggested here are extremely difficult and
require careful and detailed analysis of very high energy deep inelastic
scattering experiments; some of them might even be a prohibitive task.
However, we think that the possible outcome -- a better knowledge of the
intimate nucleon structure -- would certainly justify a serious
consideration of such an analysis.

There is another consideration which increases the phenomenological
interest of such measurements: some of the quantities we have dealt
with are not affected by the axial anomaly.
A general treatment of the perturbative gluonic contributions
can be found in Ref.\vog\ and, for charged current DIS, in Ref.\gluon.
In the case of neutral current, only the parity
violating spin structure functions, the ones which measure the
``valence''  C-odd spin contribution $\Delta q- \Delta\bar q$,
like $g_3$ and $g_5$, appear to be independent of the axial anomaly. From
a rigorous point of view, beyond the scope of this paper,
only these structure functions are unambiguously defined,
scale--independent, and describe well--defined quark spin degrees of
freedom. The comparison between these C-odd spin structure functions
and the C-even ones could be the only experimental test of whether the
anomaly is a large or small
$x$ effect \ref\bass{S. D. Bass and A. W. Thomas,
{\it J. Phys.} {\bf G19} (1993) 925}.
\listrefs
\bye